\documentclass{aa}

\usepackage{amssymb,amsmath}
\usepackage{graphicx}
\usepackage{natbib}
\usepackage{bm}
\usepackage{booktabs}
\usepackage{url}
\usepackage{xcolor}

\bibpunct{(}{)}{;}{a}{}{,} % to follow the A&A style

\def\gcm{g~cm$^{-3}$}
\def\fm{fm$^{-3}$}

\def\Zav{\langle Z \rangle}

\def\beq{\begin{equation}}
\def\eeq{\end{equation}}
\def\beqn{\begin{eqnarray}}
\def\eeqn{\end{eqnarray}}

\def\j{^{(j)}}

\begin{document}

\title{Inner crust of a neutron star at crystallization in a multi-component 
  approach
\thanks{{The tables of the impurity parameter at the crystallization temperature shown in Fig.~\ref{fig:qimp_icrust}, 
as well as tables of the impurity parameter for different values of density and temperature relevant for the inner crust, are available at the CDS via anonymous ftp to cdsarc.u-strasbg.fr (130.79.128.5) or via 
http://cdsweb.u-strasbg.fr/cgi-bin/qcat?J/A+A/}}
}
\titlerunning{Inner crust of a neutron star in a MCP}
\authorrunning{Carreau et al.}

\author{T. Carreau\inst{1},  
%N. Chamel\inst{2}, 
A.~F. Fantina\inst{2},
F. Gulminelli\inst{1}
%J.~M. Pearson\inst{4}
%et al.
}
\institute{LPC (CNRS/ENSICAEN/Universit\'e de Caen Normandie), UMR6534, 14050 
  Caen C\'edex, France \\
\email{carreau@lpccaen.in2p3.fr}
%\and Institut d'Astronomie et d'Astrophysique, CP-226, Universit\'e Libre de Bruxelles, 1050 Brussels, Belgium \\
\and Grand Acc\'el\'erateur National d'Ions Lourds (GANIL), CEA/DRF -
 CNRS/IN2P3, Boulevard Henri Becquerel, 14076 Caen, France \\
%\and D\'ept. de Physique, Universit\'e de Montr\'eal, Montr\'eal, Qu\'ebec H3C 3J7 Canada
}
   
\date{Received xxx Accepted xxx}

\abstract{%Context. 
The possible presence of amorphous and heterogeneous phases in the inner crust 
of a neutron star is expected to reduce the electrical conductivity of the 
crust, with potentially important consequences on the magneto-thermal evolution 
of the star. In cooling simulations, the disorder is quantified by an impurity 
parameter which is often taken as a free parameter.}
{%Aim.
We aim to give a quantitative prediction of the impurity parameter as a 
function of the density in the crust, performing microscopic calculations 
including up-to-date microphysics of the crust.
} 
{%Method.
A multi-component approach is developed at finite temperature using a 
compressible liquid drop description of the ions with an improved energy 
functional based on recent microscopic nuclear models and optimized on extended 
Thomas-Fermi calculations. Thermodynamic consistency is ensured by adding a 
rearrangement term and deviations from the linear mixing rule are included in 
the liquid phase.
}
{%Results.
The impurity parameter is consistently calculated at the crystallization 
temperature as determined in the one-component plasma approximation for the 
different functionals. Our calculations show that at the crystallization 
temperature the composition of the inner crust is dominated by nuclei with 
charge number around $Z \approx 40$, while the range of the $Z$ distribution 
varies from about $20$ near the neutron drip to about $40$ closer to the 
crust-core transition. This reflects on the behavior of the impurity parameter 
that monotonically increases with density up to around $40$ in the deeper 
regions of the inner crust.
%Our analysis also points out the importance of the rearrangement term, which is necessary to recover the correct limit at zero temperature. AFF: last sentence eventually to modify once new figure on rearrangement is added
}
{%Conclusion.
Our study shows that the contribution of impurities is non-negligible, thus 
potentially having an impact on the transport properties in the neutron-star 
crust. The obtained values of the impurity parameter represent a lower limit; 
larger values are expected in the presence of non-spherical geometries and/or 
fast cooling dynamics.
} 


\maketitle

%%%%%%%%%
\section{Introduction}\label{sect:introd}
%%%%%%%%%

It is generally assumed that the composition of an isolated neutron star (NS) 
is that of ``cold catalyzed matter'', i.e.~that determined in the ground state 
at zero temperature (see, e.g., \citet{hpy2007, lrr, Blaschke2018}).
In this hypothesis, the crust of a NS is supposed to be made of pure layers, 
each consisting of a one-component Coulomb crystal (except possibly at the 
interface between different layers where multinary compounds may exist; 
see \citet{chf2016a} for a discussion).
However NSs, being born from core-collapse supernova explosions, are initially 
hot, with temperatures exceeding $10^{10}$~K.
At such temperatures, the crust of a (proto-)NS is expected to be made of a 
Coulomb liquid composed of different nuclear species in a charge compensating 
electron background, see \citet{oertel2017} for a review.
As the NS crust cools down, it is generally supposed that this multi-component 
plasma (MCP) remains in full thermodynamic equilibrium until the ground state 
is reached.
However, it is unlikely that full equilibrium is maintained, after 
crystallization occurs, until $T=0$~K.
Moreover, if the NS cools down rapidly enough, the composition of the crust 
could be frozen at some finite temperature $T_{\rm f}$ above the 
crystallization temperature $T_{\rm m}$ (see, e.g., \citet{goriely2011}).
Therefore, a more realistic picture of the crust would be that of a 
multi-component solid.

For the outer crust, the co-existence of different nuclear species is not 
expected to significantly impact the static properties of the crust.
Indeed, because of the relatively low crystallization temperatures, 
$T_{\rm m} \lesssim 10^9$~K, the most probable nucleus is very close to the 
ground-state one-component plasma (OCP) composition, and the contribution of 
other ions is typically very small.
In the inner crust, the situation is a priori less obvious and deviations from 
the ground-state composition may be larger, due to the larger crystallization 
temperature, $10^9 \lesssim T_{\rm m} \lesssim 10^{10}$~K \citep{hpy2007}.
Indeed, \citet{jones1999, jones2001} suggested that thermal fluctuation of the 
charge and neutron numbers may be quite significant for mass densities 
$\rho_B \gtrsim 10^{13}$~\gcm~near the crystallization temperature. 
The presence of amorphous and heterogeneous phases in the inner crust leads to 
a higher temperature-independent electrical resistivity and strong ohmic 
dissipation, and important consequences on the magnetic field evolution were 
predicted by \citet{jones2004}. More recently, \citet{Pons2013} suggested that 
the increased resistivity due to the amorphous structure could reflect into 
observational timing properties of x-ray pulsars.
More generally, the presence of impurities in the crust has notable effects on 
transport and magneto-rotational properties of the NS (see, e.g., 
\citet{SchSht2018, GouEsp2018} for recent reviews), which, in turn, affect the 
NS thermal evolution.
For these reasons, although cooling simulations are usually carried out using 
the ground-state composition, the presence of various nuclear species is taken 
into account via an ``impurity factor'', often taken as a free parameter 
adjusted on observational cooling data, see for instance \citet{Vigano2013}. 

A microscopic calculation of the impurity parameter at the crystallization 
temperature for the outer crust of a non-accreting unmagnetized NS has been 
recently performed in \citet{fantina2020}.
In the latter work, the nuclear distributions of the multi-component liquid 
plasma at crystallization has been computed fully self-consistently, adapting a 
general formalism originally developed for the description of supernova cores 
\citep{gulrad2015, grams2018}. 
The crystallization temperature was determined in the OCP approximation, using 
a microscopic nuclear mass model based on deformed Hartree–Fock–Bogoliubov 
calculations (HFB-24, \citet{goriely2013}). 
The study of \citet{fantina2020} performed on the outer crust has been 
subsequently extended in \citet{Carreau2020}, who calculated the 
crystallization temperature and the associated composition in the inner crust 
using the compressible liquid-drop (CLD) model approach of \citet{Carreau2019}, 
with parameters optimized on four different microscopic models, namely BSk22, 
BSk24, BSk25, and BSk26, developed by \citet{goriely2013}. Shell effects, as 
calculated in \citet{pearson2019} for the same functionals, were added to the 
CLD model.
The use of such an approach instead of a fully microscopic one not only reduces 
the computational time, but more importantly allows to quantitatively estimate 
the model dependence of the results.
The outcomes of \citet{Carreau2020} suggest that, while shell effects are 
important at the lowest densities close to the outer crust, the highest source 
of uncertainties comes from the smooth part of the nuclear functional, 
specifically the surface tension at extreme isospin values.

In the present work, we employ the same CLD model with parameters optimized on 
the same functionals as in \citet{Carreau2020}, but we extend it by including a 
nuclear distribution in a MCP approach at equilibrium similar to that of 
\citet{fantina2020}. This also allows us to calculate the impurity parameter in 
the inner crust self-consistently, thus complementing the results obtained in 
\citet{fantina2020} for the outer crust.

The formalism is described in Sect.~\ref{sect:model}.
The numerical results are presented in Sect.~\ref{sect:results}; specifically, 
the composition of the inner crust is discussed in Sect.~\ref{sect:res-comp} 
and the impurity parameter in Sect.~\ref{sect:qimp}.
Finally, we conclude in Sect.~\ref{sect:conclus}.

%%%%%%%%%
%\section{Model of the crust}
\section{Model of the inner crust}\label{sect:model}
%%%%%%%%

%%%%%%%%%%%%%%%%%%%%%%%%%%%%%%%%%%% 
\subsection{MCP in nuclear statistical equilibrium}\label{sect:mcp}
%%%%%%%%%%%%%%%%%%%%%%%%%%%%%%%%%%%

To model a full statistical equilibrium of ions in the inner crust,
we extend the formalism of \citet{fantina2020} allowing for the presence of 
dripped neutrons, which are supposed to constitute a homogeneous gas. 
The possible contribution of a free proton gas is expected to be small at the 
temperatures we considered, and it will be neglected. This working hypothesis 
is a-posteriori confirmed by the calculation of the proton fugacity, 
$z_p = \exp[(\mu_p-m_p c^2)/(k_B T)]$, $\mu_p$ ($m_p$) being the proton 
chemical potential (mass), $c$ the speed of light and $k_B$ the Boltzmann 
constant, which never exceeds $-20$ MeV in the density and temperature domain 
studied in this paper. 

The NS crust at a given depth in the star is supposed to contain different ion 
species with mass and charge number $(A\j, Z\j)$, associated to different 
Wigner-Seitz cells of volume $V\j$, such that $p_j$ is the frequency of 
occurrence or probability of the component $(j)$, with  $\sum_j p_j=1$. 
Thermodynamic quantities are defined in terms of the ion densities of the 
different species  $n_N \j$, which are related to the probabilities $p_j$ 
through $n_N\j={p_j}/{\langle V\rangle}$, where the bracket notation 
$\langle \rangle$ indicates ensemble averages.
 
The different $(A\j, Z\j)$ configurations are associated with different 
baryonic densities $n_B\j$, such that the total baryonic density is 
$n_B = \sum_j p_j n_B\j$ (see Eq.~(\ref{eq:nB})).
Conversely, they share the same total pressure $P$ imposed by the hydrostatic 
equilibrium and the same background densities of electrons, $n_e\j = n_e$, and 
of free neutrons, $n_g\j = n_g$.
We also suppose that charge neutrality is realized in each cell, meaning that 
the proton density is the same in each cell, i.e. $n_p\j = n_p$, and equal to 
the electron density $n_e$, i.e. $n_e=n_p=Z\j/V\j$.

The free energy density of the multi-component system is defined as:
\begin{equation}
  \mathcal F = \sum_j n_N\j F \j \ ,\label{eq:ftot}
\end{equation}
where the free energy per ion of the component $(j)$ accounts for the 
contribution of the ion, the dripped neutrons and the electrons,
\begin{equation}
  F\j = F_i\j + F_n \j + F_e \j \ ,
\end{equation}
including their mutual interactions\footnote{We denote with capital letters the 
(free) energy per ion, e.g. $F$, while the notation $\mathcal{F}$ is used for 
the free energy density.}. 
For future convenience, the nuclear interactions between the ion and the 
neutron gas, and the Coulomb interactions between the ion and the electrons, 
are all included in the term $F_i \j$. Therefore, the free neutron and electron 
components are simply given by:
\begin{equation}
 F_n \j= V\j \mathcal F_g \; ; \;  F_e \j= V\j \mathcal F_e \ ,
\end{equation}
where $\mathcal F_{g(e)}$ is the free energy density of a uniform neutron 
(electron) gas at density $n_g$ $(n_e)$.
The explicit expression of these terms is discussed in Sect.~\ref{sect:CLD}.
The ion contribution, $F_i\j$, can be written as: 
\begin{equation}
  F_i\j = F_i^{(j),{\rm 0}} + \delta F\j \ .\label{eq:fimcp}
\end{equation}
{
The first term in Eq.~(\ref{eq:fimcp}), $ F_i^{(j),{\rm 0}}$, noting $m_n$ ($m_p$) the neutron (proton) mass, is given by
\beqn
  F_i^{(j),{\rm 0}} &=&  (A\j-Z\j) m_n c^2 + Z\j m_p c^2 + F_i^{(j),{\rm nuc}} \nonumber \\
 % F_i^{(j),{\rm 0}} = M^{\star (j)} c^2 
  && + F_i^{(j),{\rm id}} + F_i^{(j),{\rm int}} \ ,
  \label{eq:F0}
\eeqn
%
%where the effective mass reads
%
%\begin{equation}
%  M^{\star (j)} c^2 = (A\j-Z\j) m_n c^2 + Z\j m_p c^2 + F_i^{(j),{\rm nuc}} \ ,
 % \label{eq:mstar}
%\end{equation}
%
where $F_i^{(j),{\rm nuc}}$ is the internal nuclear free energy and  $F_i^{(j),{\rm int}}$ is the Coulomb interaction contribution.
The explicit expressions of these terms, as well as of the last term in Eq.~(\ref{eq:fimcp}), $\delta F\j$, accounting for the interaction between the ion and the surrounding (neutron) gas, depend on the adopted model and will be discussed in Sects.~\ref{sect:OCP} and \ref{sect:CLD}.
}
%The explicit expressions of this latter term, as well as of the Coulomb interaction contribution, $F_i^{(j),{\rm int}}$, and of the  last term in Eq.~(\ref{eq:fimcp}), $\delta F\j$, accounting for the interaction between the ion and the surrounding (neutron) gas, depend on the adopted model and will be discussed in Sects.~\ref{sect:OCP} and \ref{sect:CLD}.}
%Finally, \fg{ $ F_i^{(j),{\rm nuc}}$ is the internal nuclear free energy,} $F_i^{(j),{\rm int}}$ is the Coulomb interaction contribution and the last term in Eq.~(\ref{eq:fimcp}), $\delta F\j$, accounts for the interaction between the ion and the surrounding (neutron) gas.
%The explicit expressions of the latter terms depend on the adopted model and will be discussed in Sects.~\ref{sect:OCP} and \ref{sect:CLD}.
Finally, since in this work we are only interested in temperatures higher or equal to the melting temperature, where the MCP is expected to be in the liquid phase, the ``ideal'' contribution, $ F_i^{(j),\rm id}$, accounts for the translational center-of-mass motion:
\begin{equation}
  F_i^{(j),\rm id} = k_\textrm{B} T 
  \left[ \ln \left( \frac{n_N\j (\lambda\j)^3}{g_s\j} \right) -1 \right] \ ,
  \label{eq:fidmcp}
\end{equation}
where $g_s\j$ is the spin degeneracy, which we take $g_s\j=1$ for nuclei whose 
ground-state angular momentum is unknown, and the de Broglie wavelength of 
component $(j)$ is given by 
\begin{equation}
  \lambda^{(j)} = \sqrt{\frac{2\pi{(\hbar c)}^2}{M^{\star(j)}c^2 k_B T}} \ ,
  % \lambda^{(j)} = \sqrt{\frac{2\pi{(\hbar c)}^2}{ k_B T F_i^{(j),{\rm 0}}  }} \ ,
  \label{eq:lambda}
\end{equation}
{
$\hbar$ being the Planck-Dirac constant, and the effective mass of the ion $M^{*(j)}$ is defined as
\begin{equation}
 M^{\star (j)} c^2 = (A\j-Z\j) m_n c^2 + Z\j m_p c^2 + F_i^{(j),{\rm nuc}} + \delta F\j \ .
 \label{eq:mstar}
 \end{equation}
}
The probabilities $p_j$ and the densities $n_N\j$ are calculated such as to maximize the thermodynamic potential in the canonical ensemble. 
Because of the chosen free energy decomposition, we can observe that the electron and free neutron part of the free energy density, $\mathcal F_e$ and $\mathcal F_g$, do not depend on $n_N\j$, i.e.
\begin{equation}
  \mathcal F\left (\left \{ n_N\j \right \}\right ) = 
  \mathcal F_i\left (\left \{ n_N\j \right \}\right ) 
  + \mathcal F_e +\mathcal F_g \ ,
\end{equation}
where 
\begin{equation}
  \label{eq:FN}
  \mathcal F_i = \sum_j n_N\j F_i \j \ .
\end{equation}
Therefore, the variation can be performed on the ion part only:
\begin{equation}
\label{eq:variation}
d\mathcal F_i
%&=& \sum_j  \left ( F_N\j + n_N\j \frac{\partial F_N^{(j)}}{\partial n_N\j }+k_\textrm{B}T\right ) dn_N\j  \nonumber \\
= \sum_j  \left ( \Omega_i\j +k_\textrm{B} T\ln n_N\j
\right ) dn_N\j  \ ,
\end{equation}
where the single-ion canonical potential is given by:
\begin{eqnarray}
\Omega_i\j &=& \left( F_i\j - F_i^{(j),{\rm id}} \right) + k_B T \ln \frac{\left( \lambda\j\right)^3}{g_s\j} \nonumber \\
 && +  n_N\j \frac{\partial \left( F_i\j - F_i^{(j),{\rm id}} \right) }{\partial n_N\j } \ .
  \label{eq:omegaj}
\end{eqnarray}
In Eq.~(\ref{eq:variation}), the variations $dn_N\j$ are not independent 
because of the normalization of probabilities, and the baryonic number and 
charge conservation laws:
\begin{eqnarray}
  \frac{1}{\langle V\rangle} &=& \sum_j n_N\j \ ,  \label{eq:alpha} \\
  n_B -n_g &=&\sum_j n_N\j A\j \left (1-\frac{n_g}{n_0^{(j)}}\right ) \ , 
  \label{eq:nB} \\
  n_p&=& \sum_j n_N\j Z\j \ . \label{eq:np}
\end{eqnarray}
The correction factor on the right hand side of Eq.~(\ref{eq:nB}) accounts for the excluded volume, i.e.~the gas cannot occupy the nucleus volume.
In the same equation, $n_B$ is the total baryonic density and $n_0^{(j)}$ is the average density of the ion $(j)$. 
This latter can be calculated by imposing equilibrium with the nucleon gas via:
\begin{equation}
    \frac{n_0^{(j)2}}{A\j}
    \frac{\partial  F_i^{(j),{\rm 0}} }{\partial n_0\j}%\bigg|_{A,I,n_p,n_g} 
= P_g \ , \label{eq:n0}
\end{equation}
where $P_g=n_g^2d({\mathcal F}_g/n_g)/dn_g$ is the pressure of the neutron gas.
This expression is explicitly demonstrated in Sect.~\ref{sect:CLD} (see Eq.~(\ref{eq:eq4})).

The constraints Eqs.~(\ref{eq:alpha})-(\ref{eq:np}) are taken into account by introducing Lagrange multipliers ($\alpha,\mu_n,\mu_p$) leading to the following equations for the equilibrium densities $n_N\j $:
\begin{eqnarray}
&& \sum_j \left ( \Omega_i\j 
  +k_\textrm{B} T\ln n_N\j -\alpha
\right ) dn_N\j   \nonumber \\
&-& \mu_n \sum_j N \j dn_N\j   %\nonumber \\ 
- \mu_p \sum_j Z\j dn_N\j=0 \ ,
\end{eqnarray}
with $N \j= A\j \left (1-n_g/n_0^{(j)}\right ) -Z\j$.
Considering independent variations, the equilibrium distributions are given by 
\begin{equation}
  \label{eq:pj}
  p_j =  {\mathcal N} 
  \exp \left (- \frac{\tilde \Omega_i^{ (j)}}{k_\textrm{B} T} \right ) \  ,
\end{equation}
with the normalization
\begin{equation}
 {\mathcal N} =\exp\left(\frac{\alpha}{k_\textrm{B}T}\right) = 
 \sum_j \exp 
 \left (- \frac{\tilde \Omega_i^{ (j)}}{k_\textrm{B} T} \right ) \  .
\end{equation}
The single-ion grand-canonical potential $ \tilde \Omega_N^{(j)}$ reads:
\begin{equation}
\tilde \Omega_i^{ (j)}=  \Omega_i\j -\mu_n N\j - \mu_p Z\j
 \  , 
 \label{eq:gnuc}
\end{equation}
where $\mu_n$ and $\mu_p$ can be identified with the neutron and proton 
chemical potentials, respectively.
In the definitions above, the ion free energy contains the rest-mass energy, thus the chemical potentials include the rest-mass energies as well.

The calculation of the grand-canonical potential, $\tilde \Omega_i^{ (j)}$, requires the evaluation of the chemical potentials $\mu_n$, $\mu_p$, as well as of the rearrangement term {(last term in Eq.~(\ref{eq:omegaj}))}, 
\begin{equation}
\mathcal R\j= n_N\j 
\frac{\partial \left( F_i\j - F_i^{(j),{\rm id}} \right)} {\partial n_N\j } \ .
\end{equation} 
These terms will be worked out in Sects.~\ref{sect:chem} and~\ref{sect:rearrangement}, respectively.

Once the abundancies of the different ions are calculated via Eq.~(\ref{eq:pj}) at the crystallization temperature, it is also possible to calculate the impurity parameter of the solid crust, which represents the variance of the ionic charge distributions and is defined as (see, e.g., the discussion in Sect.~7 in \citet{meisel2018} for a review)
\begin{equation}
  \label{eq:Qimp}
  Q_{\rm imp} =\sum_j p(Z\j) (Z\j - \Zav)^2 \ ,
\end{equation}
where $p(Z\j)$ is the normalized probability distribution (integrated over all $N\j$) of the element $Z\j$.

%%%%%%%%%%%%%%%%%%%%%%%%%%%%%%%%%%
\subsection{Evaluation of the chemical potentials}\label{sect:chem}
%%%%%%%%%%%%%%%%%%%%%%%%%%%%%%%%%%

In a given thermodynamic condition expressed by a temperature $T$ and a 
pressure $P$, the proton and neutron chemical potentials can be determined 
using the thermodynamic relation $\mathcal{F} + P = \mu_n n_n + \mu_p n_p + 
\mu_{e} n_{e}$, giving, together with the beta equilibrium condition $\mu_n = 
\mu_e + \mu_p$ ($\mu_e$ being the electron chemical potential),
\begin{equation}
  \mu_n = \frac{\mathcal{F} + P}{n_B} \; ; \; 
  \mu_e= \frac{\mathcal{F}_{e} + P_{e}}{n_p} \ ,
  \label{eq:mutrue}
\end{equation}
where the baryon and proton densities, $n_B$ and $n_p$, are given by 
Eqs.~(\ref{eq:nB}) and (\ref{eq:np}), respectively, the free energy density 
$\mathcal{F}$ is given by Eq.~(\ref{eq:ftot}), and $\mathcal{F}_{e}$ 
($P_{e}=n_e^2d({\mathcal F}_e/n_e)/dn_e$) is the free energy density (pressure) 
of the electron gas at density $n_p = n_{e}$. 
With this prescription, the equilibrium probabilities can only be determined by 
the solution of a complex non-linear system of coupled equations which is a 
challenging numerical task.  
Such a complete nuclear statistical equilibrium formalism has been adopted by 
different authors (see \citet{oertel2017,Burgio2018} for a review); however, 
simplified nuclear functionals were adopted, the density (instead of the 
pressure) was imposed, and the rearrangement term was neglected. 
 
In the outer crust regime, it was found by \citet{fantina2020} that a 
perturbative implementation of the nuclear statistical equilibrium as proposed 
by \citet{grams2018} leads to a very fast convergence, with reduced 
computational cost and increased numerical precision. 
We therefore adopt this same prescription in the inner crust.
In the perturbative treatment, the equilibrium problem is solved in the OCP 
approximation, as detailed in Sect.~\ref{sect:OCP} below. This gives a first 
guess for the chemical potentials as:
\begin{equation}
  \mu_n^{\rm OCP} = \frac{\mathcal{F^{\rm OCP}} + P}{n_B^{\rm OCP}} \; ; \; 
  \mu_p^{\rm OCP} = \mu_n^{\rm OCP} - 
  \frac{\mathcal{F}_{e} + P_{e}}{n_p^{\rm OCP}}, 
  \label{eq:muOCP}
\end{equation}
where $\mathcal{F^{\rm OCP}}$ is the equilibrium energy density in the OCP 
approximation, and $n_B^{\rm OCP}$, $n_p^{\rm OCP}$ are the baryon and proton 
densities that, in the OCP approximation, lead to the pressure $P$ 
(see Sect.~\ref{sect:OCP}).  
Similarly, the electron quantities $\mathcal{F}_{e}$ and $P_{e}$ are calculated 
at $n_e=n_p^{\rm OCP}$.
With this guess, the ion abundancies are readily calculated via 
Eq.~(\ref{eq:pj}), and using again Eq.~(\ref{eq:mutrue}) we can get an improved 
estimation of the chemical potentials as
\begin{eqnarray}
  \label{eq:mun-g}
  \mu_n &=&   \frac{\sum_j n_N\j F\j}{n_B \langle V \rangle} 
  + \frac{P}{n_B} \ ,  \label{eq:munMCP} \\
  y_p \mu_e &=&  \frac{\sum_j n_N\j F_e\j}{n_B \langle V \rangle} 
  + \frac{P_e}{n_B} \ ,\label{eq:mueMCP}
\end{eqnarray}
{where $y_p=\langle Z\rangle/(n_B \langle V \rangle)$ is the average proton 
fraction of the mixture, with $\langle Z \rangle = \sum_j p_j Z\j$.}
The problem can thus be solved by iteration. 
It turns out that the difference between the initial guess Eq.~(\ref{eq:muOCP}) 
and the result of the first iteration, 
Eqs.~(\ref{eq:munMCP})-(\ref{eq:mueMCP}), is so small for all pressures and 
temperatures considered in this work, that the simple OCP estimation, 
Eq.~(\ref{eq:muOCP}), can be kept.
 
The full MCP calculation becomes therefore computationally equivalent to the 
much simpler OCP one, with the additional advantage that the MCP results can be 
compared to the more standard OCP ones with no extra computational cost.

%%%%%%%%%%%%%%%%%%%%%%%%%
\subsection{The OCP approximation}\label{sect:OCP}
%%%%%%%%%%%%%%%%%%%%%%%%%

In the OCP approximation, the equilibrium configuration of inhomogeneous dense 
matter in the inner crust in full thermodynamic equilibrium is obtained by 
minimizing the free-energy density in a Wigner-Seitz cell of volume $V$ with 
the constraint of a given baryon density $n_B$, 
see \citet{lattimer1991,gulrad2015,Carreau2020}.

Similarly to the general MCP case of Sect.~\ref{sect:mcp} we write:
\begin{equation}
\mathcal{F}(A,I,n_0,n_p,n_g)
=\frac{F_{i}+F_n+F_e}{V} , \label{eq:focp}
\end{equation}
where the variational variables are the mass number $A$ and isospin ratio 
$I=1-2Z/A$ of the ion, its internal density $n_0$, the proton density in the 
cell $n_p=n_e$, and the density of the homogeneous gas of dripped neutrons 
$n_g$.  
As in Sect.~\ref{sect:mcp}, see Eq.~(\ref{eq:fimcp}), we include the 
interactions of the nucleus with the neutrons and electrons in the term $F_i$:
\begin{equation}
  F_i = F_i^{{\rm 0}}  + \delta F \ ,
  \label{eq:fiocp}
\end{equation}
{
with
  \beqn
    F_i^{{\rm 0}} &=& (A-Z) m_n c^2 + Z m_p c^2 + F_i^{\rm nuc} \nonumber \\
    && +  F_i^{\rm id}+  F_i^{\rm int} \ .
    \label{eq:finucocp}
  \eeqn
  }
  %
%  with
  %
%  \begin{equation}
 %   M^{\star} c^2 = (A-Z) m_n c^2 + Z m_p c^2 +  F_i^{\rm nuc} \ . 
 %   \label{eq:mstarocp}
%  \end{equation}
In the OCP approximation, the translational motion is limited to the single Wigner-Seitz cell ($n_N = 1/V$):
\begin{equation}
 F_i^{\rm id}=k_B T \, 
 \left[ \ln \left( \frac{ \lambda^3}{V g_s} \right) - 1 \right]  \, 
 \label{eq:fidocp}
\end{equation}
where the de Broglie wavelength $\lambda$ is given by {the same expression as in Eq.~(\ref{eq:lambda}), with $M^{\star (j)}=M^{\star}$}.
The interacting part of the ion free energy can be decomposed as:
\begin{equation}
  \label{eq:Fint}
  F_i^{\rm int}=F_{ii, {\rm liq}} + F_{ie, {\rm liq}}^{\rm pol} \ .
\end{equation}
{
Analytical formulae have been derived by \citet{pc2000} for these two terms; see their Eqs.~(16) and (19), respectively.
For this study, only the first term is included; indeed, the polarization correction is found to have no effect in the density and temperature regime studied in the present paper and is therefore neglected. 
In addition, the nuclear finite-size correction is also included.
The latter is derived from the Gauss theorem and reads
\beq
E_{\rm fs} = \frac{2n_p}{n_0(1-I)}\frac{e^2}{r_0}\frac{Z^2}{A^{1/3}} \ ,
\label{eq:efinsize}
\eeq
with  $r_0 = (4\pi n_0/3)^{-1/3}$.
}

Finally, the interaction between the ion and the surrounding neutron gas is treated in the excluded volume approximation:
\begin{equation}
  \delta F = - \frac{A}{n_o} \mathcal{F}_g \ .
  \label{eq:deltaF}
\end{equation}

The equilibrium configuration is obtained by minimizing Eq.~(\ref{eq:focp}) 
with respect to the variational variables using the baryon density constraint 
limited to a single cell,
\begin{equation}
  n_B=n_g+\frac{A}{V}\left (1-\frac{n_g}{n_0} \right ).
\end{equation}
This leads to the following system of coupled differential equations\footnote{
These equations are equivalent to Eqs.~(8)-(11) in \citet{Carreau2020}. Indeed, 
the notation $F_i$ in \citet{Carreau2020} is equivalent to the notation 
$F_i^{\rm 0}$ used in the present paper. We note however that there is a 
misprint in Eq.~(9) in \citet{Carreau2020} (although the calculations have been 
done correctly); indeed, the term $\Delta m_{n,p} c^2$ should not appear in 
their Eq.~(9).}:
\begin{eqnarray}
    \frac{\partial (F_{i}^{\rm 0}/A)}{\partial A}%\bigg|_{I,n_0,n_p,n_g}
& = &0, \label{eq:eq1} \\
    \frac{2}{A}\left ( \frac{\partial F_{i}^{\rm 0}}{\partial I}%\bigg|_{A,n_0,n_p,n_g} 
- \frac{n_p}{1-I}  \frac{\partial F_{i}^{\rm 0}}{\partial n_p}%\bigg|_{A,I,n_0,n_g}
\right )  &=& \mu_{e} , \label{eq:eq2} \\
    \frac{F_{i}^{\rm 0}}{A} 
    + \frac{1-I}{A}\frac{\partial F_{i}^{\rm 0}}{\partial I}%\bigg|_{A,n_0,n_p,n_g} 
&=& \mu_B - \frac{P_g}{n_0}, \label{eq:eq3} \\
    {n_0}^2\frac{\partial (F_{i}^{\rm 0}/A)}{\partial n_0}%\bigg|_{A,I,n_p,n_g} 
&=& P_g, \label{eq:eq4}
\end{eqnarray}
where the gas pressure is given by $P_g=n_g^2d({\mathcal F}_g/n_g)/dn_g=n_g\mu_B-\mathcal{F}_g$, and the baryon 
chemical potential $\mu_B$ results:
\begin{equation}
  \mu_B =  \frac{2 n_pn_0}{n_0(1-I) - 2 n_p}
  \frac{\partial (F_{i}^{\rm 0}/A)}{\partial n_g}%\bigg|_{A,I,n_0} 
  + \frac{d\mathcal{F}_g}{dn_g} .
\end{equation}
In our parametrization (see Sect.~\ref{sect:CLD}), the in-medium modification 
of the nuclear energy arising from the external gas is governed by a single 
parameter $p$ which does not depend on the external neutron density but only on 
the isospin asymmetry $I$. 
Therefore, $\partial F_{i}^{\rm 0}/\partial n_g=0$ and the baryon chemical 
potential can be identified with the chemical potential of the gas, 
$\mu_B = \mu_g\equiv  d \mathcal{F}_g/ d n_g$.

At each value of the baryon density $n_B^{\rm OCP}$ and temperature 
$T\ge T_{\rm m}$ above the crystallization point, the system of coupled 
differential equations, Eqs.~(\ref{eq:eq1})-(\ref{eq:eq4}), is numerically 
solved as in~\cite{Carreau2019}.
This procedure leads to the determination of the favored liquid composition 
$(A,I,n_0,n_p,n_g)|_{\rm OCP}$ and to the evaluation of the total free energy 
and pressure, $\mathcal{F}^{\rm OCP}$ and $P$, as well as of the electron 
component, $\mathcal{F}_e(n_e^{\rm OCP})$, $P_e(n_e^{\rm OCP})$. 
These quantities allow one to compute the chemical potentials of the MCP using 
Eq.~(\ref{eq:muOCP}).

 %%%%%%%%%%%%%%%%%%%%%%%%%%%%%%%
\subsection{The free energy functional}\label{sect:CLD}
%%%%%%%%%%%%%%%%%%%%%%%%%%%%%%%

The free energy functional for an isolated nucleus in the vacuum is modeled 
using the CLD model of \citet{Carreau2020}, that we briefly recall.
 
The nuclear free energy $F_i^{{\rm nuc}}$ at temperature $T$ of a nucleus of 
mass number $A$, isospin asymmetry $I$, and average density $n_0$, is 
decomposed into a bulk, surface and Coulomb part as:
\begin{equation}
    F_i^{\rm nuc}= Af_{b}(n_0,I,T)   + F_{\rm {surf+curv}} + F_{\rm Coul}, 
    \label{eq:fnuc}
\end{equation}
where $f_b(n_B,\delta,T)$ represents the free energy per baryon of bulk nuclear 
matter, with $n_B=n_p+n_n$, $\delta=(n_n-n_p)/n_B$, and $n_p (n_n)$ is the 
homogeneous proton (neutron) density.  
Assuming spherical nuclei, we write the Coulomb energy as:
\begin{equation}
  F_{\rm Coul}= \frac{3}{5} \frac{e^2}{r_0} \frac{Z^2}{A^{1/3}} \ .
\end{equation}
with % $r_0 = (4\pi n_0/3)^{-1/3}$, 
$e$ the elementary charge, and the surface and curvature free energies as in~\citet{Newton2013,lattimer1991}:
\begin{eqnarray}
   \label{eq:Fsc}
    F_{\rm {surf+curv}} & = & 4\pi r_0^2\sigma_s A^{2/3} \nonumber \\
    & & + 8\pi r_0\sigma_s\frac{\sigma_{0,c}}{\sigma_0}
    \alpha\left(\beta-\frac{1-I}{2}\right)A^{1/3}, 
\end{eqnarray}
with $\alpha = 5.5$, and an isospin dependent surface tension given by: 
\begin{equation}
    \sigma_s = \sigma_0 
    \frac{2^{p+1} + b_s}{(Z/A)^{-p} + b_s + (1-Z/A)^{-p}}. \label{eq:sigma}
\end{equation}
The surface and curvature parameters $\sigma_0$, $b_s$, $p$, $\sigma_{0,c}$, 
and $\beta$ are optimized on extended Thomas-Fermi (ETF) mass tables built with 
the same nuclear functional adopted for the bulk term, see \citet{Carreau2020} 
for details. 
The same functional is also used to compute the free energy density of the 
neutron gas,
 \begin{equation}
\mathcal{F}_g=n_g f_b(n_g,1,T) + n_g m_n c^2 \ .
\end{equation}
In principle, a shell and pairing correction should be added to the nuclear 
free energy expression, Eq.~(\ref{eq:fnuc}). 
However, it was shown in \citet{Carreau2020} that these corrections rapidly 
fade away with the temperature, and we will thus consider that they can be 
neglected in the temperature range we explore in this work.

Concerning the nuclear models, we use the same functionals as in 
\citet{Carreau2020}, namely the recent functionals of the BSk family BSk22, 
BSk24, BSk25, and BSk26 introduced by \citet{goriely2013}. 
These realistic microscopic models span a relative large range in the symmetry 
energy parameters consistent with existing experimental constraints thus 
covering the most important part of the present EoS uncertainty 
\citep{pearson2014, pearson2018}, meaning that the spread of the predictions of 
those models can be taken as a reasonable estimation of the model dependence of 
our results.

{
Finally, the free energy density $\mathcal{F}_{e}$ and pressure $P_e$ of the electron gas are calculated within a relativistic Sommerfeld expansion.
The complete expressions can be found in \citet{hpy2007}, see their Eqs.~(2.65) and (2.67), respectively. 
Exchange and correlation contributions are found to be very small in the ranges of density and temperature explored in this work and can be safely neglected.
}

%%%%%%%%%%%%%%%%%%%%%%%%%%
\subsection{Evaluation of the rearrangement term}\label{sect:rearrangement}
%%%%%%%%%%%%%%%%%%%%%%%%%%

The computation of the equilibrium distributions, Eq.~(\ref{eq:pj}), associated 
to a thermodynamic condition characterized by a temperature $T$ and chemical 
potentials $\mu_n, \mu_p$, requires the evaluation of the rearrangement term 
entering Eq.~(\ref{eq:omegaj}),
\begin{equation}
  \mathcal R\j= n_N\j 
  \frac{\partial\left( F_i\j - F_i^{(j),{\rm id}} \right) }{\partial n_N\j} \ .
\end{equation}

As already discussed in \citet{fantina2020}, the rearrangement term arises from 
the self-consistency induced by the Coulomb part of the ion free energy.
This stems from the fact that, due to the strong incompressibility of the 
electrons, we have imposed charge conservation at the  level of each cell:  
\begin{equation}
  n_e=n_p = \sum_j n_N\j Z\j = \frac{Z\j}{V\j}   \ .
  \label{eq:conscharge} 
\end{equation}
This is at variance with the baryonic density that can fluctuate from cell to 
cell, see Eq.~(\ref{eq:nB}). 
As a consequence of that, any component of the free energy density that depends 
on the local cell proton density $n_p\j=n_p$ leads to a dependence on the local 
density $n_N\j$ through Eq.~(\ref{eq:conscharge}). 
Within the functional described in Sect.~\ref{sect:CLD}, this is only the case 
for the Coulomb interaction $F_i^{(j),{\rm int}}$. 
The rearrangement term of component $(j)$ thus reduces to:
\begin{eqnarray}
\mathcal{R}\j &=& n_N\j \left. 
  \frac{\partial F_i^{(j),{\rm int}}}{\partial n_N\j} 
  \right|_{\{n_N^{(i)}\}_{ i\ne j}} \nonumber \\
                    &=&  n_N^{(j)}  Z^{(j)} \frac{\partial
                    F_i^{(j),{\rm int}}}{\partial n_p} \ ,
 \label{eq:rear}
\end{eqnarray}
where we have used Eq.~(\ref{eq:conscharge}) implying the relation 
$\partial n_p / \partial n_N\j = Z\j$.
Following~\cite{grams2018}, to avoid the complication of a self-consistent 
resolution of Eq.~(\ref{eq:pj}), we look for an approximation of 
Eq.~(\ref{eq:rear}) using the requirement that the most probable ion in the MCP 
mixture should coincide with the OCP result, if non-linear mixing terms in the 
MCP are omitted. This condition is a direct consequence of the principle of 
ensemble equivalence in the thermodynamic limit, see \citet{gulrad2015}.

To look for the extremum of Eq.~(\ref{eq:pj}), one has to consider that in the 
MCP both $n_p$ and $n_g$ are imposed once the thermodynamic condition is 
specified. Therefore, these densities do not act anymore as constraints and 
should not be varied. The variation of Eq.~(\ref{eq:pj}) with respect to the 
ion variables $A$, $I$, $n_0$ thus gives:
\begin{eqnarray}
  \frac{{n_0}^2}{A} \left(\frac{\partial  F_i^{\rm 0}}{\partial n_0} 
    + \frac{\partial
  \mathcal{R}}{\partial n_0}\right) &=& P_g,\label{eq:eqn0} \\
  \frac{2}{A}\left(\frac{\partial  F_i^{\rm 0}}{\partial I} + \frac{\partial
  \mathcal{R}}{\partial I}\right) &=& \mu_n -
  \mu_p , \label{eq:eqaa2} \\
  \frac{\partial  F_i^{\rm 0}}{\partial A} 
  + \frac{\partial \mathcal{R}}{\partial A} +
  \frac{1-I}{A}\left(\frac{\partial F_i^{\rm 0}}{\partial I} +
\frac{\mathcal{R}}{\partial I}\right) & =& \mu_n -
\frac{P_g}{n_0} , \label{eq:eqii2}
\end{eqnarray}
where the partial derivatives are calculated at the values corresponding to the 
equilibrium OCP solution and $F_i^{\rm 0}$ is given by Eq.~(\ref{eq:finucocp}) 
using Eq.~(\ref{eq:fidocp}), i.e.~by the OCP functional, that is non-linear 
mixing terms are excluded.

By comparing Eqs.~(\ref{eq:eqn0})-(\ref{eq:eqii2}) to the OCP ones, 
Eqs.~(\ref{eq:eq1})-(\ref{eq:eq4}), and using $P_g=n_g\mu_n-{\mathcal F}_g$, we 
can deduce that ${\mathcal R}\j$ should not depend on $n_0\j$, 
i.e. $\mathcal R\j=\mathcal R\j(A\j, I\j)$, and that at the OCP solution we 
should have:
\begin{equation}
  \frac{1-I}{A}\frac{\partial \mathcal R}{\partial I} 
  = -  \frac{\partial \mathcal R}{\partial A} .
\end{equation}
This is satisfied if ${\mathcal R}\j$ linearly depends on $Z\j=A\j(1-I\j)/2$. 
Our final expression for the rearrangement term is therefore:
\begin{equation}
  \mathcal{R}^{(j)} \simeq  Z^{(j)}\left\langle
    \frac{\langle n_N^{(j)} \rangle \partial F_i^{(j),{\rm int}}}{\partial n_p}
  \right\rangle_j
  = Z^{(j)} \left(\frac{1}{V}\frac{\partial
  F_i^{\rm int}}{\partial n_p}\right)_{\rm OCP},
\end{equation}
where the quantity in the parenthesis is calculated at the OCP solution.

%%%%%%%%%
\section{Numerical results}\label{sect:results}
%%%%%%%%

We computed the finite-temperature composition of the inner crust of 
non-accreting unmagnetized NSs within our MCP approach, thus including a 
distribution of nuclei in nuclear statistical equilibrium. For the considered 
BSk functionals, the recent calculations of \citet{Pearson2020} show that 
non-spherical pasta structures are expected to be present at the highest 
densities above $n_B\approx 0.05$ fm$^{-3}$, close to the crust-core transition 
point. Since we only consider spherical nuclei in the present study, we limited 
our calculation to the density domain extending from the neutron drip point to 
$n_B= 0.04$ fm$^{-3}$.
 
All the results presented in this Section were obtained using the BSk CLD 
models, with the surface and curvature parameters fitted to the corresponding 
Extended Thomas-Fermi calculations and  crust-core transition densities (see 
Table 1 of \citet{Carreau2020} for the explicit parameter values).
In Sect.~\ref{sect:res-comp}, the results for the inner-crust composition at 
finite temperature are shown for the BSk24 CLD model, as an illustrative 
example, while the impurity parameter is presented in Sect.~\ref{sect:qimp} for 
all the four considered CLD models based on the BSk22, BSk24, BSk25, and BSk26 
functionals.

Our calculations of the liquid MCP are performed at the crystallization 
temperature $T_{\rm m}$ and, for comparison, at $10^{10}$~K = T > $T_{\rm m}$.
The reason of this choice stems from the fact that, depending on the NS cooling 
timescales, the composition may be already frozen at some temperature 
$T_{\rm f} > T_{\rm m}$ (see e.g. \citet{goriely2011}).
In \citet{Carreau2020}, the crystallization temperature of the inner crust was 
estimated to lie between $\approx 2.5 \times 10^9$~K and $\approx 8 \times 
10^9$~K for the considered CLD models (see their Figs.~5 and 7, panel (a)).
Therefore, we have chosen $T_{\rm f} = 10^{10}$~K as an illustrative example.
Indeed, a more realistic estimate of $T_{\rm f}$ would require dynamical 
simulations, which are beyond the scope of this paper.

\subsection{Equilibrium composition of the MCP}\label{sect:res-comp}

\begin{figure}[htbp]
  \begin{center}
    \includegraphics[width=\linewidth]{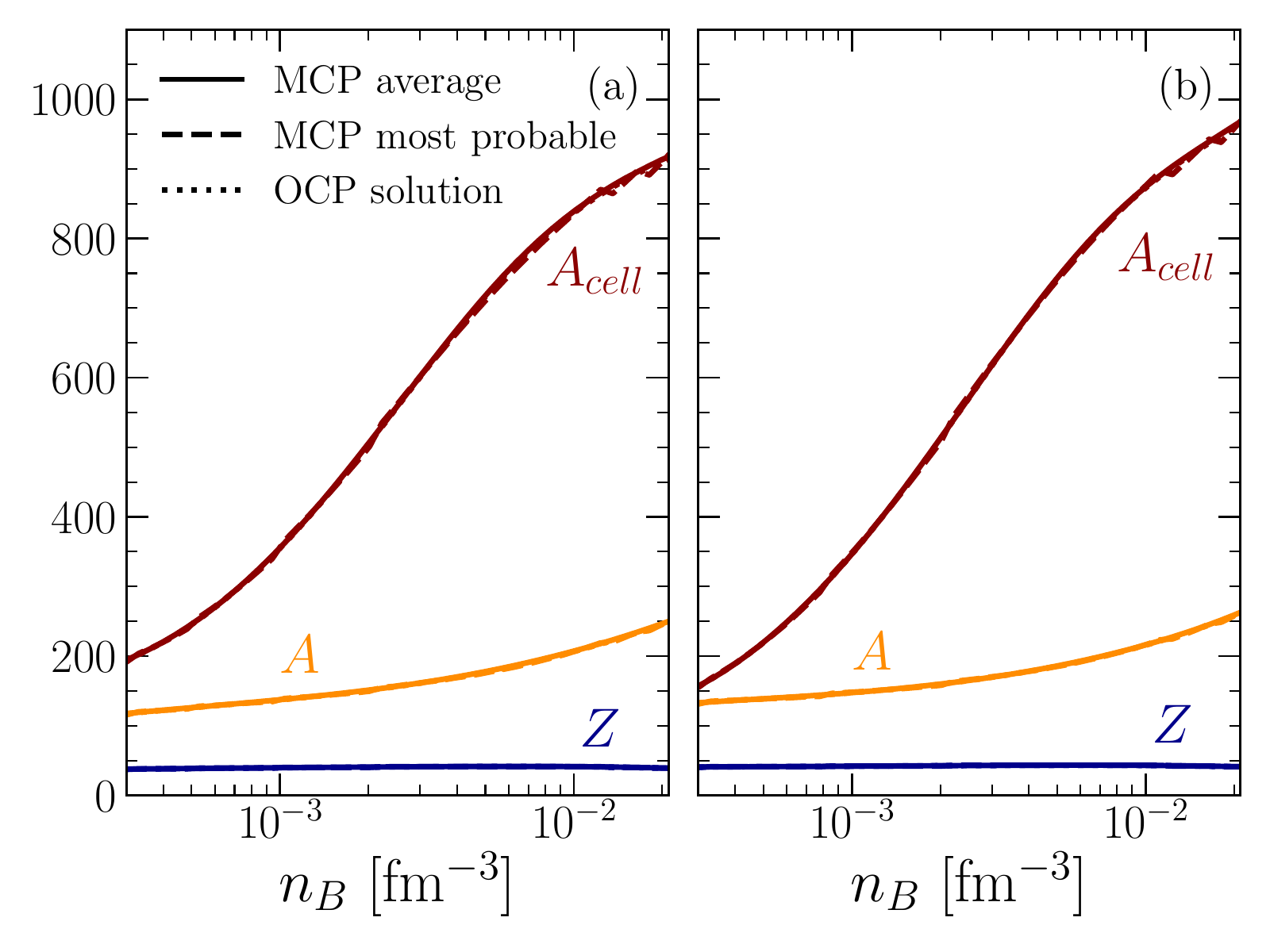}
  \end{center}
  \caption{Variation with baryon density $n_B$ of the average (solid lines) 
    and most probable (dashed lines) values of the charge number $Z$ 
    (blue lines), cluster mass number $A$ (orange lines) and total mass number 
    $A_{\rm cell}$ (red lines) in the inner crust at two selected 
    temperatures: $T=10^{10}$ K (panel (a)), and $T=T_{\rm m}$ 
    (panel (b)). Results obtained in the one-component plasma (OCP) approximation are also shown 
  (dotted lines).
}
\label{fig:compo_icrust_bsk24}
\end{figure}

The average and most probable mass and charge number in the MCP are displayed 
in Fig.~\ref{fig:compo_icrust_bsk24} as a function of the baryon density in the 
inner crust for $T = 10^{10}$~K (panel (a)) and $T = T_{\rm m}$ (panel (b)).
For comparison, the results obtained in the OCP approximation are also shown 
(dotted lines).
We can see that the average and most probable values in the MCP approach follow 
very closely the OCP ones.
This means that the deviations from the linear mixing rule in the liquid phase 
are small, as already noticed in \citet{fantina2020} for the outer crust. 
While the mass numbers increase with density, the charge number is almost 
constant, $Z \approx 40$. The latter value is very close to that obtained at 
zero temperature (see also the dotted curve in Fig.~6, panel (b), in 
\citet{Carreau2020} and Fig.~12 in~\cite{pearson2018}), suggesting that the 
presence of $Z \approx 40$ ions in the inner crust is a robust result.

\begin{figure}[htbp]
  \begin{center}
    \includegraphics[width=\linewidth]{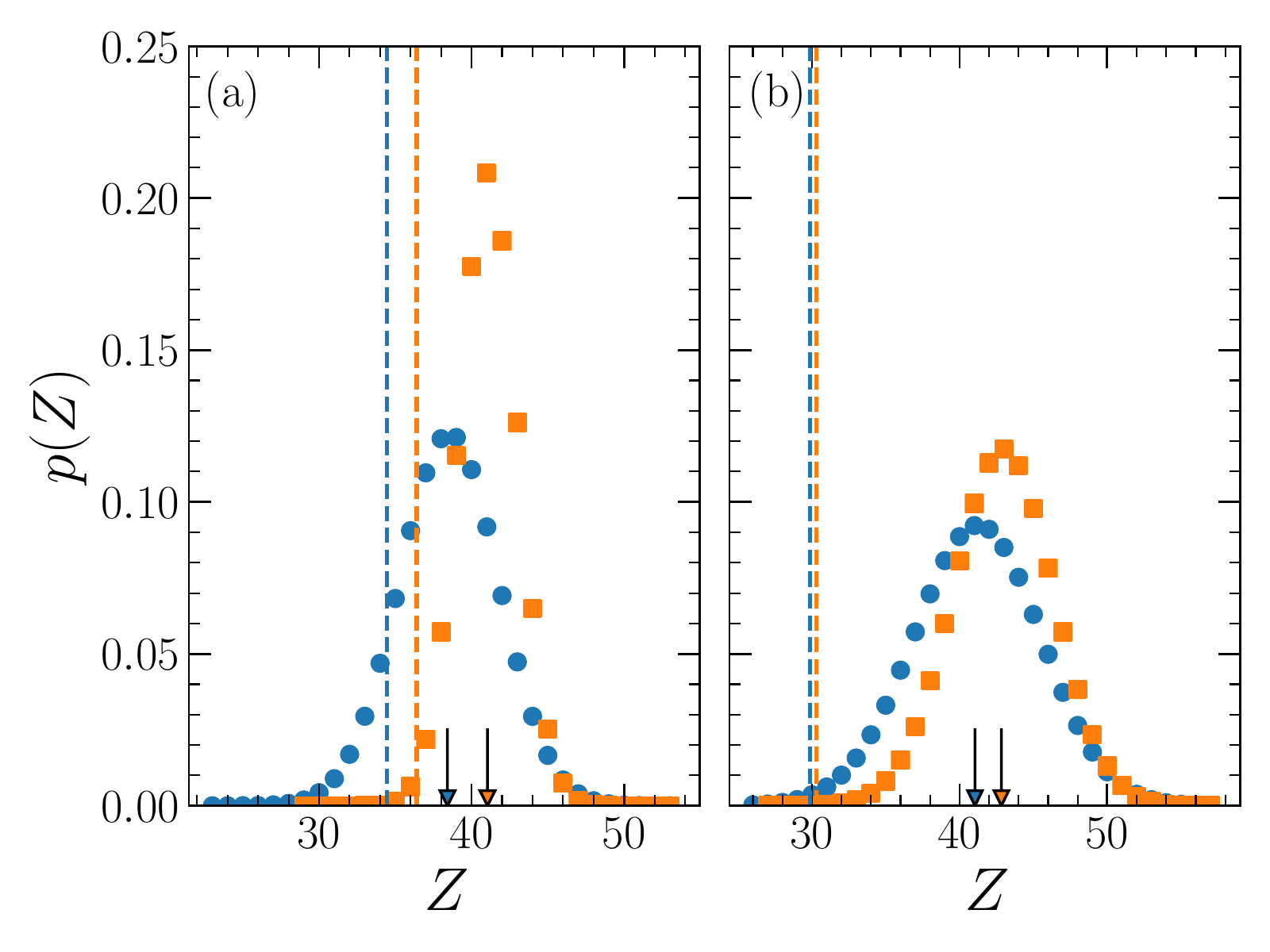}
  \end{center}
  \caption{Normalized probability distribution $p(Z)$ for $n_B = 5 \times 
    10^{-4}$ fm$^{-3}$ (panel (a)) and $n_B = 10^{-2}$ fm$^{-3}$ (panel (b)) at 
    two selected temperatures: $T=10^{10}$ K (orange squares), and 
    $T=T_{\rm m}$ (blue circles). Arrows indicate the OCP solutions.
    Vertical dashed lines correspond to the value of $\langle Z \rangle$ obtained 
    without considering the rearrangement term (see text for details).
  }
\label{fig:pj_icrust_bsk24}
\end{figure}

\begin{figure}[htbp]
  \begin{center}
    \includegraphics[width=\linewidth]{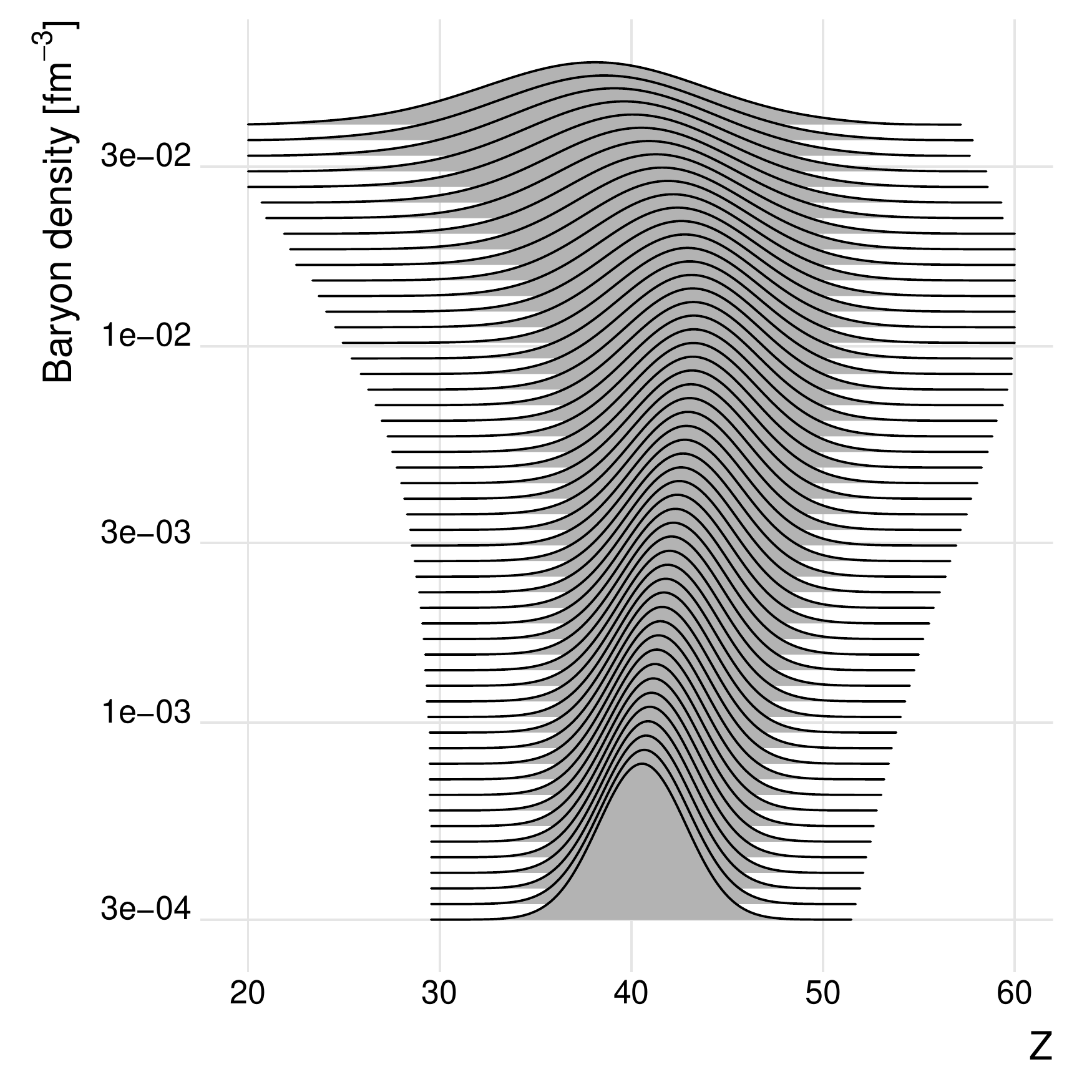}
  \end{center}
  \caption{Normalized probability distribution $p(Z)$ with increasing 
    baryon density $n_B$ in the inner crust at the crystallization
    temperature $T_{\rm m}$. 
    }\label{fig:joyplot_bsk24_tm}
\end{figure}

To evaluate the width of the distribution, we show in Fig.~\ref{fig:pj_icrust_bsk24} the normalized probability distribution $p(Z)$ 
for $T = 10^{10}$~K and $T = T_{\rm m}$ and for two selected densities in the inner crust: $n_B = 5 \times 10^{-4}$~\fm~(panel (a)) and $n_B = 10^{-2}$~\fm~(panel (b)). 
{
The peaks of the distributions, i.e.~the most probable $Z$, coincide with the charge numbers predicted in the OCP approximation (shown by the associated arrows), thus indicating that the linear mixing rule is a good approximation. 
To assess the importance of the rearrangement term, Eq.~(\ref{eq:rear}), we mark with vertical lines the average values of the charge number, $\langle Z \rangle$, obtained when this term is not included in the calculations. 
We observe that the effect of the rearrangement term is significant, particularly at higher density.
Indeed, without taking into account this term, the distribution is systematically and considerably shifted towards lower $Z$, proving that the rearrangement term is actually needed to satisfy the thermodynamic consistency.
}
We can also notice that, as expected, the distributions become broader with increasing 
temperature and density, thus making the OCP approximation less reliable. 
The flattening of the distribution is more clearly visible in Fig.~\ref{fig:joyplot_bsk24_tm} where the normalized probability distribution $p(Z)$ at the crystallization temperature is plotted for different increasing baryon densities in the inner crust. 
While the average value of $Z$ is centered around $40$ throughout the 
inner crust, the range of $Z$ of the distribution varies from $\approx 20$ 
closer to the neutron drip up to $\approx 40$ near the crust-core transition.

\begin{figure}[htbp]
  \begin{center}
    \includegraphics[width=\linewidth]{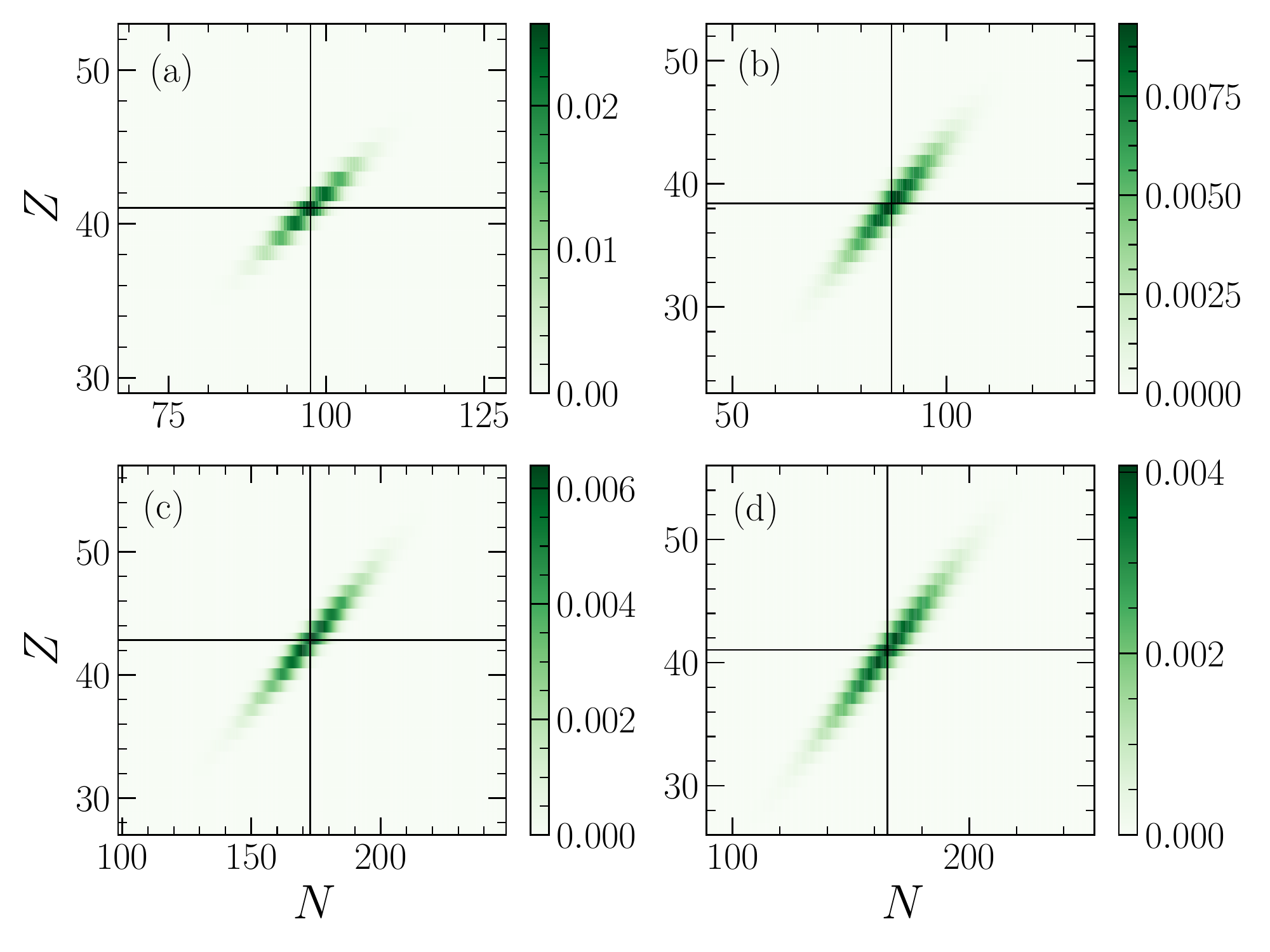}
  \end{center}
  \caption{Normalized probability distribution of nuclei $p(N,Z)$ for four 
    chosen thermodynamic conditions. 
  Panel (a): $n_B = 5\times 10^{-4}$~fm$^{-3}$, $T=T_{\rm m}$; 
  Panel (b): $n_B = 5\times 10^{-4}$~fm$^{-3}$, $T=10^{10}$~K; 
  Panel (c): $n_B = 10^{-2}$~fm$^{-3}$, $T=T_{\rm m}$; 
  Panel (d): $n_B = 10^{-2}$~fm$^{-3}$, $T=10^{10}$~K. 
  In each panel, the OCP solution coincides with the intersection of the black lines.
  }\label{fig:aa_zz_bsk24}
\end{figure}

To better assess the evolution of the nuclear distribution, both in charge and 
mass number with density and temperature, we show in Fig.~\ref{fig:aa_zz_bsk24} 
the normalized probability distribution $p(Z,N)$ for two selected densities in 
the inner crust: $n_B = 5 \times 10^{-4}$~\fm~(panels (a) and (b)) and $n_B = 
10^{-2}$~\fm~(panels (c) and (d)), both at $T = 10^{10}$~K (panels b and d) and 
$T = T_{\rm m}$ (panels (a) and (c)). As expected, going from lower to higher 
densities (upper to lower panels), we observe that the ions species become more 
neutron rich and that the distribution, both in $Z$ and $N$, broadens when 
going from lower to higher temperatures (left to right panels).

\subsection{Impurity parameter}\label{sect:qimp}

\begin{figure}[htbp]
  \begin{center}
    \includegraphics[width=\linewidth]{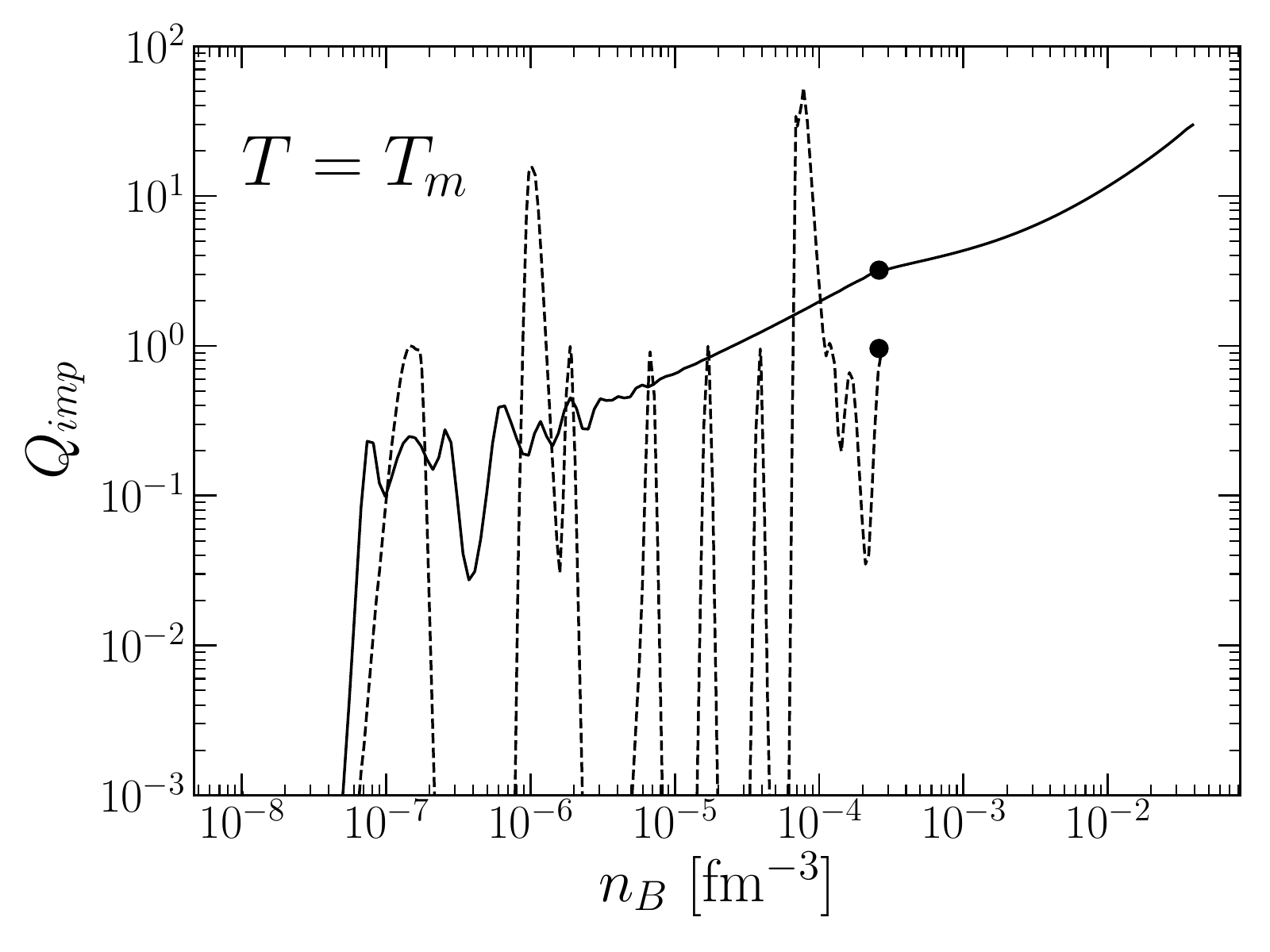}
  \end{center}
  \caption{Variation with baryon density $n_B$ of the impurity parameter 
    $Q_{\rm imp}$ in the crust at the crystallization temperature 
    $T=T_{\rm m}$. In the outer crust regime, solid line (dashed line) 
    represents the BSk24 CLD (HFB-24) prediction. In the inner crust regime, 
    the impurity parameter is calculated in the CLD approximation. Points 
    indicate the neutron-drip transition.
  }
  \label{fig:qimp_crust_bsk24}
\end{figure}

The impurity parameter at the crystallization temperature is shown in the whole 
crust in Fig.~\ref{fig:qimp_crust_bsk24}, as calculated with the BSk24 CLD 
model (solid line). The black dot marks the neutron drip point. 
We can see that the impurity parameter 
in the inner crust is higher than in the outer crust, meaning that the 
distribution is less peaked thus the OCP approximation is less reliable than in 
the outer crust.
Indeed, larger values of $Q_{\rm imp}$ indicate more appreciable deviations 
from the OCP predictions. For comparison, we also plot the impurity parameter, 
taken from \citet{fantina2020}, calculated in the outer crust with the HFB-24 
model (dashed line). The latter calculations show more prominent variations of 
$Q_{\rm imp}$ with respect to the CLD model calculations. This is due to the 
natural inclusion of shell effects in the fully microscopic calculations, that 
exhibit bimodal distributions around values of pressure, corresponding to the 
simultaneous presence of the two characteristic elements of adjacent layers 
(see Fig.~6 in \citet{fantina2020}). 
These strong fluctuations are naturally smoothed out in the CLD model, because 
the nuclear functional varies continuously with $A$ and $Z$, and so does the 
probability. However, we can observe that the CLD calculation nicely 
interpolates the microscopic results, with an average impurity factor steadily 
increasing with the density and lying in the interval 
$Q_{\rm imp}\approx 0,1-2$. 
Going in the inner crust, neutrons drip out of the finite ion volume, and the 
associated  shell effects naturally disappear \citep{chamel2006}. 
The inclusion of proton shell effects in the inner crust would require a 
formidable numerical investment which is much beyond the purpose of the paper. 
Moreover, it was suggested in \citet{Carreau2020} that these effects are small 
at the higher melting temperature of the inner crust, and that their effect on 
the observables is smaller than the uncertainty brought by our imperfect 
knowledge of the smooth part of the energy functional. 
For this reason shell effects were completely neglected in our study. 
We expect that a fully microscopic calculation would still present oscillations 
in $Q_{\rm imp}$ beyond the drip point, that these oscillations should be 
progressively damped going deeper in the star, and that our calculation can be 
taken as a smooth interpolation of that oscillating behavior.

\begin{figure}[htbp]
  \begin{center}
    \includegraphics[width=\linewidth]{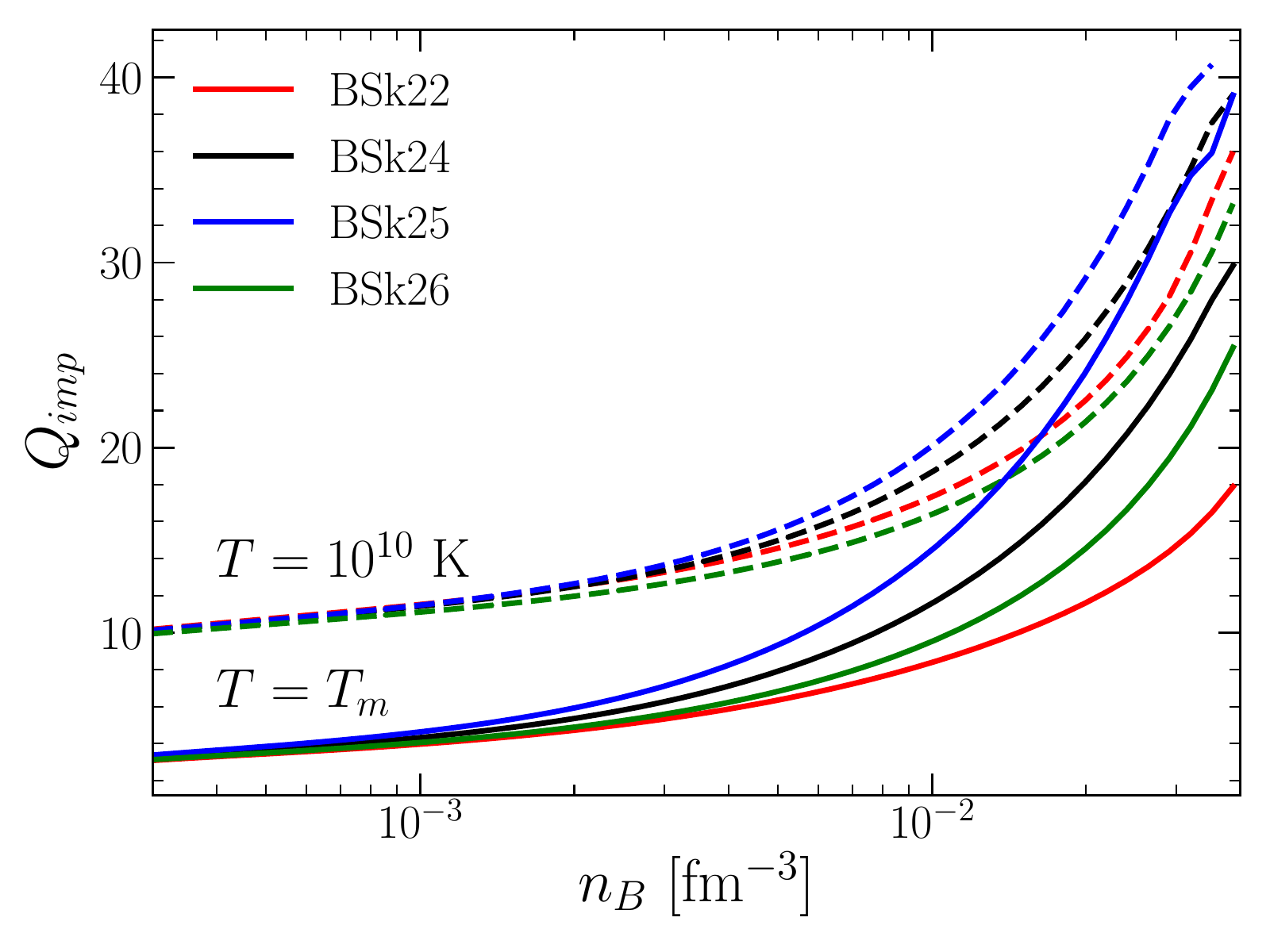}
  \end{center}
  \caption{Variation with baryon density $n_B$ of the impurity parameter 
    $Q_{\rm imp}$ in the inner crust regime at two selected temperatures: 
    $T=10^{10}$~K, and $T=T_{\rm m}$, based on BSk22 (red lines), BSk24 (black 
    lines), BSk25 (blue lines), and BSk26 (green lines) CLD calculations. 
    %with $p=3$.
  }\label{fig:qimp_icrust}
\end{figure}

To have a quantitative prediction of the impurity factor, the problem of model 
dependence has to be addressed. Apart from the modeling of finite temperature 
shell effects discussed above, the main source of uncertainty of the 
calculation comes from the choice of the nuclear functional. 
We show, in Fig.~\ref{fig:qimp_icrust}, the impurity parameter, 
Eq.~(\ref{eq:Qimp}), as a function of the baryon density in the inner crust, at 
the crystallization temperature $T_{\rm m}$ (solid line), for the four 
considered BSk CLD models.
{
These data, as well as tables of the impurity parameter on a density-temperature grid for the four considered models, are available in tabular format at the CDS.
}
Considering that the chosen models are believed to cover the main uncertainty 
on the nuclear equation of state at sub-saturation density \citep{pearson2018}, 
we can take the spread of $Q_{\rm imp}$ values as obtained by the four 
calculations, as a reasonable estimation of the uncertainty on the impurity 
parameter.
Since this latter represents the variance of the charge distribution, low values of $Q_{\rm imp}$ indicate that the distribution is quite peaked and thus that the OCP approach is a good approximation, as it can also be seen from Figs.~\ref{fig:compo_icrust_bsk24} and~\ref{fig:pj_icrust_bsk24}, panel (a).
The monotonic increase of the impurity parameter with density is also in accordance with Fig.~\ref{fig:joyplot_bsk24_tm}, which clearly show the growth of the width of the charge distribution with increasing density.
While at lower densities all the models predict similar values of the impurity parameter ($\lesssim 5$), at higher densities the spread among the models becomes larger, with the model associated to the lower (larger) symmetry energy coefficient at saturation having the larger (lower) $Q_{\rm imp}$.
The same trend is observed at $T=10^{10}$~K (dashed lines in Fig.~\ref{fig:qimp_icrust}), although the hierarchy of the models is not preserved. 

%%%%%%%%%
\section{Conclusions}\label{sect:conclus}
%%%%%%%%

In this paper, we have presented a multi-component approach for the modeling of 
the crust of isolated unmagnetized NSs.
Completing the work of \citet{fantina2020} on the outer crust, the composition 
of the inner crust is evaluated with an extended nuclear statistical 
equilibrium based on a CLD model for the nuclear part of the ion energetics.
To achieve thermodynamical consistency, a rearrangement term is explicitly 
worked out. This term has an important effect on the distributions and it is 
necessary to recover the correct limit at zero temperature.
Since NSs are born hot, the equilibrium composition of a mature NS can be 
determined assuming a liquid phase for the MCP, at the lowest temperature at 
which strong and weak equilibrium are attained. In the absence of a dynamical 
estimation of the associated reaction rates, we consider the lowest temperature 
limit as given by the OCP crystallization temperature $T_{\rm m}$.
We show that at that temperature the OCP approximation gives a very good 
estimation of the average composition of the inner crust, non-linear mixing 
terms playing a very small role in the liquid phase.
However, an important contribution of impurities is obtained, favoring the 
picture of a temperature-independent high resistivity in the inner crust for 
all $T<T_{\rm m}$.

In order to reach quantitative predictions for the associated impurity 
parameter, we consider four different realistic microscopic nuclear functionals 
of the BSk family, which cover the present uncertainty in the nuclear modeling 
below the nuclear matter saturation density.   
The impurity parameter is seen to increase with the density, and values in the 
interval $Q_{\rm imp}\approx 20-40$ are reached at the highest densities 
considered in this study, namely $n_B=0.04$ fm$^{-3}$.  
Higher values of the impurity parameter might be expected in the deepest region 
of the inner crust, close to the core-crust transition, due to the presence of 
non-spherical pasta phases, which have not been considered in the present 
study.

%%% ACKOWLEDGEMENTS
\begin{acknowledgements}
This work has been partially supported by the IN2P3 Master Project MAC and the 
CNRS PICS07889. Discussions with N.~Chamel are gratefully acknowledged.
%and the PHAROS European Cooperation in Science and Technology (COST) action CA16214.
%The work of N.C. was supported by Fonds de la Recherche Scientifique (Belgium) under grant IISN 4.4502.19.
\end{acknowledgements}

%%%%%%%%%%%%%%%%%%%%%%%
%%% BIBLIOGRAPHY

%%%%%%%%%%%%%%%%%%%%%%%%%%%%%%%%%%%%%%%%%%%%%%%%%%%%%%%%%%%%%%


\newpage
\vfill
\begin{thebibliography}{}
%{\small

%\bibitem[Audi et al.~(2012)]{ame2012} Audi, G., Wang, M., Wapstra, A.~H., Kondev, F.~G., MacCormick, M., Xu, X., \& Pfeiffer, B.\ 2012, Chinese Physics C, 36, 1287
%
%\bibitem[Baym et al.~(1971a)]{bps} Baym, G., Pethick, C., \& Sutherland, P. \ 1971, Astrophys. J., 170, 299
%\bibitem[Baym et al.~(1971b)]{bbp} Baym, G.~A., Bethe, H.~A., \& Pethick, C.~J.\ 1971, Nucl. Phys. A, 15, 225
%
%\bibitem[Baiko et al.~(2001)]{baiko2001} Baiko, D.~A., Potekhin, A.~Y., \& Yakovlev, D.~G.\ 2001, Phys.~Rev.~E, 64, 057402
%
%\bibitem[Bisnovatyi-Kogan \& Chechetkin~(1979)]{bisno1979} Bisnovatyi-Kogan, G.~S., Chechetkin, V.~M.\ 1979, 
%%“Nonequilibrium shells of neutron stars and their role in sustaining X-ray emission and nucleosynthesis,” 
%Sov. Phys. Uspekhi 22, 89–108
%%NICO new ref
\bibitem[Blaschke \& Chamel (2018)]{Blaschke2018} Blaschke, D., \& Chamel, N. \ 2018,  Astrophys. Space Sci. Libr., 457, 337-400
%
\bibitem[Burgio \& Fantina (2018)]{Burgio2018} Burgio, G.~F., \& Fantina, A.~F.\ 2018, Astrophys. Space Sci. Libr., 457, 255
%
%\bibitem[Carreau et al.~(2018)]{Carreau2019a} Carreau, T., Gulminelli, F., \& Margueron, J.\ 2019, Phys. Rev. C, 100, 055803
\bibitem[Carreau et al.~(2019)]{Carreau2019} Carreau, T., Gulminelli, F., \& Margueron, J.\ 2019, Eur. Phys. J. A, 55, 188
\bibitem[Carreau et al.~(2020)]{Carreau2020} Carreau, T., Gulminelli, F., Chamel, N., Fantina, A.~F., \& Pearson, J.~M.\ 2020, A\&A, 635, A84
%
\bibitem[Chamel~(2006)]{chamel2006} Chamel, N.\ 2006, Nucl. Phys. A, 773, 263.
\bibitem[Chamel \& Fantina~(2016)]{chf2016a} Chamel, N., \& Fantina, A. F. \ 2016, Phys. Rev. C, 94, 065802
%\bibitem[Chamel \& Fantina~(2016b)]{chf2016b} Chamel, N., \& Fantina, A. F. \ 2016b, Phys. Rev. D, 93, 063001
%\bibitem[Chamel et al. (2015)]{chamel2015} Chamel, N., Fantina, A. F., Zdunik, J.- L., Haensel, P. \ 2015, Phys. Rev. C, 91, 055803
\bibitem[Chamel \& Haensel~(2008)]{lrr} Chamel, N., \& Haensel, P.\ 2008, ``Physics of Neutron Star Crusts'', Living Reviews in Relativity 11, 10
%\bibitem[Chamel et al.~(2007)]{chamel2007} Chamel, N., Naimi, S., Khan, E., \& Margueron, J.\ 2007, Phys. Rev C, 75, 055806
%\bibitem[Douchin and Haensel~(2001)]{Douchin2001} Douchin, F., \&  Haensel, P. \ 2001,  A\&A,  380, 151
%
%\bibitem[Ducoin et al.~(2007)]{ducoin2007} Ducoin, C., Chomaz, Ph., \& Gulminelli, F.\ 2007, Nucl. Phys. A, 789, 403
%
\bibitem[Fantina et al.~(2020)]{fantina2020} Fantina, A.~F., De Ridder, S., Chamel, N., \& Gulminelli, F.\ 2020, A\&A, 633, A149
%
%\bibitem[Farouki \& Hamaguchi~(1993)]{farham1993} Farouki, R.~T., \& Hamaguchi, S.\ 1993, Phys.~Rev.~E, 47, 4330
%
%{\bibitem[Gnedin et al.~(2001)]{Gnedin2001} Gnedin, O. Y., Yakovlev, D. G., \& Pothekin, A. Y. \ 2001, Mon. Not. R. Astron. Soc. 324, 725
%}
\bibitem[Goriely et al.~(2013)]{goriely2013} Goriely, S., Chamel, N., \& Pearson, J.~M.\ 2013, Phys.~Rev.~C, 88, 024308
\bibitem[Goriely et al.~(2011)]{goriely2011} Goriely, S., Chamel, N., Janka, H.-T., \& Pearson, J.~M. \ 2011, A\&A, 531, A78
%
%{
\bibitem[Gourgouliatos \& Esposito~(2018)]{GouEsp2018} Gourgouliatos, K.~N., \& Esposito, P., in ``The Physics and Astrophysics of Neutron Stars'', edited by L. Rezzolla, P. Pizzochero, D. I. Jones, N. Rea, and I. Vida\~{n}a, Astrophysics and Space Science Library, Vol. 457, p. 57-93 (Springer, Berlin, 2018)
%}
\bibitem[Grams et al.~(2018)]{grams2018} Grams, G., Giraud, S., Fantina, A.~F., \& Gulminelli, F.\ 2018, Phys. Rev. C, 97, 035807
\bibitem[Gulminelli \& Raduta~(2015)]{gulrad2015} Gulminelli, F. \& Raduta, Ad.~R.\ 2015, Phys.~Rev.~C, 92, 055803
%
%\bibitem[Haensel and Pichon~(1994)]{hp1994} Haensel, P., \& Pichon, B.\ 1994, A\&A, 283, 313
\bibitem[Haensel et al.~(2007)]{hpy2007} Haensel, P., Potekhin, A.~Y., \& Yakovlev, D.G.\ 2007, ``Neutron Stars 1. Equation of state and structure'' (Springer, New York, 2007)
%
\bibitem[Jones~(1999)]{jones1999} Jones, P.~B.\ 1999, Phys. Rev. Lett., 83, 3589
\bibitem[Jones~(2001)]{jones2001} Jones, P.~B.\ 2001, Mon. Not. Royal Astron. Soc., 321, 167
\bibitem[Jones~(2004)]{jones2004} Jones, P.~B.\ 1999, Phys. Rev. Lett., 93, 221101
%
\bibitem[Lattimer \& Swesty~(1991)]{lattimer1991}J. Lattimer, M., \& Swesty, F.D., \ 1991, Nucl. Phys. A, 535, 331
%%\bibitem[Margueron et al.~(2018a)]{Margueron2018a} Margueron, J., Hoffmann Casali, R., \& Gulminelli, F.\ 2018a, Phys. Rev. C, 97, 025805
%%\bibitem[Margueron et al.~(2018b)]{Margueron2018b} Margueron, J., Hoffmann Casali, R., \& Gulminelli, F.\ 2018b, Phys. Rev. C, 97, 025806
%{
%\bibitem[Martin \& Urban~(2015)]{Martin2015} Martin, N., \& Urban, M., \ 2015, Phys. Rev. C, 92, 015803
%}
\bibitem[Meisel et al.~(2018)]{meisel2018} Meisel, Z., Deibel, A., Keek, L., Shternin, P., \& Elfritz, J.\ 2018, J. Phys. G, 45, 093001
%\bibitem[Negele \& Vautherin~(1973)]{Negele1973} Negele, J.W., Vautherin, D.\ 1973, Nucl. Phys. A, 207, 298–320
%
\bibitem[Newton et al.~(2013)]{Newton2013} Newton, W.~G., Gearheart, M., \& Li, B.~A.\ 2013, %A survey of the parameter space of the compressible liquid drop model as applied to the neutron star inner crust. 
%\apjs, 204, 9
%
\bibitem[Oertel et al.~(2017)]{oertel2017} Oertel, M., Hempel, M., Klähn, T., \& Typel, S., \ 2017, Rev. Mod. Phys., 89, 015007 
%\bibitem[Onsi et al.~(2008)]{onsi2008} Onsi, M., Dutta, A.~K., Chatri, H., Goriely, S., Chamel, N., \& Pearson, J.~M.\ 2008, Phys. Rev. C, 77, 065805
%
%\bibitem[Pearson et al.~(2011)]{pearson2011} Pearson, J.~M., Goriely, S., \& Chamel, N.\ 2011, Phys.~Rev.~C, 83, 065810
%\bibitem[Pearson et al.~(2012)]{pearson2012} Pearson ,J. M., Chamel, N., Goriely, S., Ducoin C. \ 2012, Phys.~Rev.~C, 85, 065803
\bibitem[Pearson et al.~(2014)]{pearson2014} Pearson, J. M., Chamel, N., Fantina, A.~F., \& Goriely, S.\ 2014, Eur. Phys. J A, 50, 43
\bibitem[Pearson et al.~(2018)]{pearson2018} Pearson, J.~M., Chamel, N., Potekhin, A.~Y., Fantina, A.~F., Ducoin, C., Dutta, A.~K., \& Goriely, S.\ 2018, Mon. Not. Royal Astron. Soc., 481, 2994
\bibitem[Pearson et al.~(2019)]{pearson2019} Pearson, J.~M., Chamel, N., Potekhin, A.~Y., Fantina, A.~F., Ducoin, C., Dutta, A.~K., \& Goriely, S.\ 2019, Mon. Not. Royal Astron. Soc., 488, 768
%{
\bibitem[Pearson et al.~(2020)]{Pearson2020} Pearson, J.~M., Chamel, N.,
Potekhin, A.~Y.\ 2020, Phys. Rev. C 101, 015802
\bibitem[Pons et al.~(2013)]{Pons2013} Pons, J. A., Viganò, D. \&  Rea, N. \ 2013, Nature Physics 9, 431.
\bibitem[Potekhin \& Chabrier~(2000)]{pc2000} Potekhin, A.~Y., \& Chabrier, G.\ 2000, Phys.~Rev.~E, 62, 8554
%\bibitem[Potekhin \& Chabrier~(2010)]{pc2010} Potekhin, A.~Y., \& Chabrier, G.\ 2010, Contrib.~Plasma~Phys.~E, 50, 82xw
%
%\bibitem[Ravenhall et al.~(1983)]{Ravenhall1983} Ravenhall, D.~G., Pethick, C.~H., \& Lattimer, J.~M.\ 1983, Nucl. Phys. A, 407, 571
%
%\bibitem[Salpeter~(1961)]{sal61} Salpeter, E.~E.\ 1961, Astrophys. J., 134, 669
%{
\bibitem[Schmitt \& Shternin~(2018)]{SchSht2018} Schmitt, A., \& Shternin, P., in ``The Physics and Astrophysics of Neutron Stars'', edited by L. Rezzolla, P. Pizzochero, D. I. Jones, N. Rea, and I. Vida\~{n}a, Astrophysics and Space Science Library, Vol. 457, p. 455-574 (Springer, Berlin, 2018)
%}
%\bibitem[Stolzmann \& Bl{\"o}cker~(1996)]{stol1996} Stolzmann, W., \& Bl{\"o}cker, T.\ 1996, A\&A, 314, 1024
%
%\bibitem[Tondeur (1971)]{tondeur1971} Tondeur, F. \ 1971, A\&A, 14, 451
\bibitem[Vigan\`o et al.~(2013)]{Vigano2013} Viganò, D., Rea, N., Pons, J. A., Perna, R., Aguilera, D. N., \& Miralles, J. A. \ 2013, Mon. Not. R. Astron. Soc. 434, 121 
%{
%\bibitem [Vi\~{n}as et al.~(2017)]{Vinas2017}  Vi\~{n}as, X., Gonzalez-Boquera, C., Sharma, B. K., \&Centelles, M.,
%\ 2017, Acta Physica Polonica B, Proceedings Supplement, 10, 259}
%\bibitem[Wang et al. (2017)]{ame2016} Wang M., Audi G., Kondev F. G., Huang W. J., Naimi S., Xu X. \ 2017, Chinese Phys. C, 41, 030003
%
%\bibitem[Weiss et al.~(2004)]{coxgiuli} Weiss, A., Hillebrandt, W., Thomas, H.-C., \& Ritter, H. ``Cox and Giuli’s Principles of Stellar Structure'', extended 2nd ed. (Cambridge Scientific Publishers, Cambridge, 2004)
%
%\bibitem[Xu et al.~(2013)]{xu2013} Xu, Y., Goriely, S., Jorissen A., Chen, G. L., \& Arnould, M.\ 2013, A\& A, 549, A106

\end{thebibliography}
\end{document}